\documentclass[aps,prb,superscriptaddress,twocolumn,showpacs]{revtex4-1}

\usepackage{amsmath, amsthm, amssymb}
\usepackage{graphicx}
\usepackage{color}

\usepackage{natbib}
\usepackage{hyperref}
\hypersetup{
        colorlinks=true,
}

\newcommand{\eqnref}[1]{(\ref{#1})}

\usepackage{tikz}
\usetikzlibrary{calc}
\usetikzlibrary{backgrounds}
\usetikzlibrary{arrows}
\usetikzlibrary{shapes.arrows}
\usetikzlibrary{decorations.markings}

\begin{document}

\title{Effect of thermal fluctuations in topological $p$-wave superconductors}

\author{Bela Bauer}
\affiliation{Station Q, Microsoft Research, Santa Barbara, CA 93106-6105, USA}

\author{Roman M. Lutchyn}
\affiliation{Station Q, Microsoft Research, Santa Barbara, CA 93106-6105, USA}

\author{Matthew B. Hastings}
\affiliation{Station Q, Microsoft Research, Santa Barbara, CA 93106-6105, USA}
\affiliation{Duke University, Department of Physics, Durham, NC 27708, USA}

\author{Matthias Troyer}
\affiliation{Theoretische Physik, ETH Zurich, 8093 Zurich, Switzerland}

\begin{abstract}
We study the effect of thermal fluctuations on the topological stability of chiral $p$-wave superconductors. We consider two models of superconductors: spinless and spinful with a focus on topological properties and Majorana zero-energy modes. We show that proliferation of  vortex-antivortex pairs above the Kosterlitz-Thouless temperature $T_{\rm KT}$ drives the transition from a thermal Quantum Hall insulator to a thermal metal/insulator, and dramatically modifies the ground-state degeneracy splitting. Therefore, in order to utilize 2D chiral $p$-wave superconductors for topological quantum computing, the temperature should be much smaller than $T_{\rm KT}$. Within the spinful chiral $p$-wave model, we also investigate the interplay between half-quantum vortices carrying Majorana zero-energy modes and full-quantum vortices having trivial topological charge, and discuss topological properties of half-quantum vortices in the background of proliferating full-quantum vortices.
\end{abstract}


\maketitle

\section{Introduction}

Topological phases of matter have been subject of intense physics research in the last decade.\cite{Wilczek09, *Nayak10, *Franz10, *Stern10} In addition to interest from the fundamental physics perspective, these states of matter can also be used for topological quantum computation,\cite{tqc} which is predicted to have an exceptional fault-tolerance by virtue of encoding and manipulating information in non-local degrees of freedom of topologically ordered systems.\cite{Kitaev97, *Freedman98, *Nayak08} Candidate physical systems include Fractional Quantum Hall states\cite{Moore_NPB91} and topological superconductors.\cite{read2000,Fu08,*Sau10}
%The common underlying features of all these systems is a ground-state degeneracy as well as the presence of certain quasiparticle excitations (non-Abelian anyons) whose manipulation allows one to process quantum information.
In all these systems the topological degrees of freedom coexist with non-topological ones. It is important to understand their interplay because it often determines the stability of the topological phase.

In this Letter we focus on 2D topological $p$-wave superconductors and study their robustness against thermal fluctuations. Specifically, we investigate the topological degeneracy in these systems in the presence of thermally-generated topological defects (vortices). Without vortices, the stability condition for the topological superconducting phase is set by the quasiparticle energy gap $\Delta$, {\it i.e.}, $T\ll T_c\sim \Delta$.\cite{cheng11} We will show that vortex-antivortex proliferation provides a more stringent temperature requirement.

We first focus on a spinless model, where we model thermal fluctuations with a classical XY model. Increasing the temperature above the Kosterlitz-Thouless (KT) transition point but still well below the local quasiparticle gap ($T_{\rm KT}< T \ll T_c$), vortices start to proliferate and eventually destroy the topological spinless superconducting phase by driving the system into a thermal metal or non-topological insulator phase.
The degeneracy splitting in the low-temperature ($T<T_{\rm KT}$) and high-temperature ($T>T_{\rm KT}$) phases changes from an exponential to a power-law scaling in the system size.

A spinful model allows for both half-quantum (HQV) and full-quantum (FQV) vortices. Only the former carry robust Majorana zero-energy modes. Thus, one can consider the interesting situation where the superconducting phase is disordered due to the presence of FQVs and study the splitting of a degeneracy due to HQVs embedded in the system at large enough separation $R$. Naively, one might expect that the splitting would not be affected by the proliferation of FQVs since the splitting energy is governed by the local quasiparticle gap which is only weakly affected by thermal fluctuations. However, we show that the situation is much more intricate and requires a deeper understanding of the interplay between topological and non-topological degrees of freedom.

Disorder in superconductors in symmetry class $D$\cite{altland1997} can drive a transition from the thermal Quantum Hall (TQH) phase to either a thermal metal (TM) phase or a topologically trivial thermal insulator (TTI).\cite{read2000, bocquet2000, Vishveshwara2000, Gruzberg2001, chalker2001, mildenberger2007} These phases have been primarily studied within network models\cite{Gruzberg2001, chalker2001, mildenberger2007} and very recently in certain microscopic models.\cite{medvedyeva2010,laumann2012} We find that all these phases appear in our microscopic model with thermally generated disorder. The TTI is an Anderson insulator with a non-zero density of states at $E=0$. We demonstrate that it does not realize the aforementioned scenario where the splitting is governed by the local quasiparticle gap if the phase is disordered. Instead, the presence of additional emergent zero-energy modes changes the splitting of the topological degeneracy from exponential to power-law.\cite{footnote1}

\begin{figure}
  \includegraphics[width=3in]{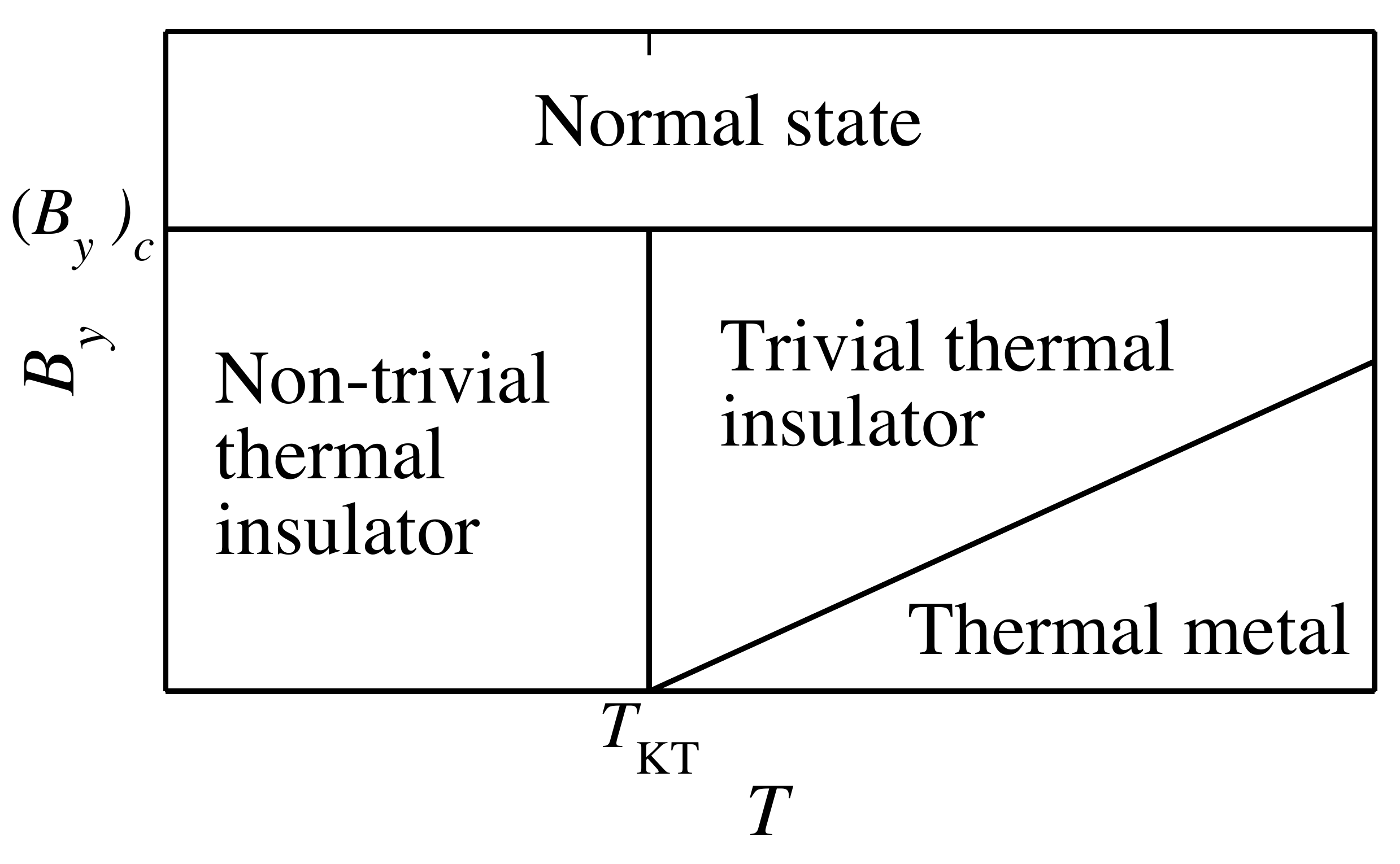}
  \caption{Schematic phase diagram of a spinful $p$-wave superconductor as a function of temperature and in-plane magnetic field. The phase diagram of the spinless case is recovered for $B_y=0$.  Above $(B_y)_c$, the superconducting order parameter vanishes in a self-consistent calculation.  \label{fig:pd} }
\end{figure}

\section{Model}

We set up our numerical problem in two steps. First, we model thermal fluctuations of the superconducting phase using a classical XY Hamiltonian. We then make an adiabatic approximation assuming that the vortex dynamics are slow compared to the quasiparticle one which appears to be quite reasonable since Abrikosov vortices are macroscopic objects and have large effective mass.~\cite{kopnin1991} Under these conditions, quasiparticles are moving in a static background of different vortex configurations. Secondly, we diagonalize the Bogoliubov-de-Gennes Hamiltonian for each disorder realization and compute the quasiparticle energy spectrum, the density of states (DoS) and inverse participation ratio (IPR). %A similar approach was used in Ref.~\cite{mayr2005}.

We consider a model for a $p+ip$ superconductor on a torus of $L \times L$ sites defined as
\begin{align}
H=&\sum_{\langle {i,j} \rangle, \sigma, \sigma'} \left(t^{\sigma \sigma'} c_{i\sigma}^\dagger c_{j\sigma'}+\Delta_{ij}^{\sigma \sigma'} c_{i\sigma}^\dagger c_{j\sigma'}^\dagger + h.c.\right) \\&-\mu \sum_{i\sigma} c_{i\sigma}^\dagger c_{i\sigma}. \nonumber
\end{align}
We first study a spinless model by choosing $t^{\sigma \sigma'}=-t\delta_{\sigma, \sigma'}$, $\Delta_{ij}^{\sigma \sigma'} = \Delta_0 \chi_{ij} \theta_{ij} \delta_{\sigma,\sigma'}$, where $\mu$ is the chemical potential. This corresponds to a $\hat d$-vector characterizing spin-triplet pairing to be aligned along $\hat x$-axis.\cite{ivanov2001}
Without spin-mixing perturbations we can equivalently study a spin-polarized system.
$\chi_{ij}$ is a chirality factor that implements $p+ip$ pairing and is $\pm 1$ for $j = i\pm \hat{x}$, and $\pm i$ for $j = i \pm \hat{y}$. $\theta_{ij}$ is a phase variable to be discussed below, and $\Delta_0$ is chosen to be a constant. We will discuss the self-consistency condition for $\Delta_0$ below. We solve the corresponding BdG equation numerically to obtain eigenvalues $E_n$ and eigenstates $(u_n, v_n)^T$.

In the homogeneous case $\theta_{ij} = 1$, the dispersion of the spinless Hamiltonian is
\begin{align}
\epsilon(k) =&   -\mu - 2t (\cos(k_x) + \cos(k_y)) \\
E(k) =& \sqrt{\epsilon(k)^2 + \Delta_0 (\sin^2(k_x) + \sin^2(k_y)) }.
\end{align}
The dispersion has gapless points for $\mu = -4t, 0, 4t$. The homogeneous system is in a topological phase if $\Delta_0 > 0$ and $-4t < \mu < 0$ or $0 < \mu < 4t$.

Vortices in a spinless $p$-wave superconductor bind Majorana zero-energy modes.\cite{read2000} These localized quasiparticles are described by a self-conjugate operator $\gamma = \gamma^\dagger$. The ground state degeneracy and Majorana quasiparticles lead to non-Abelian braiding statistics in these many-particle systems.\cite{ivanov2001,stern2004,read2009,bonderson2011} Depending on parameters, vortices may carry a large number of localized states below the bulk gap. To reduce the required computational effort, we typically use $\Delta \sim \mu$. For such a choice of parameters, the coherence length $\chi$ becomes comparable to the lattice spacing and no midgap states except the zero-energy state are present. The effect of such midgap states has been treated in Ref.~\onlinecite{meng2012} and will be discussed at the end of our paper.

The situation becomes more subtle when many vortices are present and localized zero-energy modes hybridize leading to a ground-state degeneracy splitting.~\cite{meng2009, Baraban_PRL09, Bonderson_PRL09, Lahtinen2011} At large vortex separation $R \gg \xi$, Majorana modes acquire an exponentially small energy splitting and the ground-state degeneracy at small vortex density is preserved up to exponentially small corrections $\delta E$, which for $p$-wave superconductors read\cite{meng2009, meng2010}
\begin{align} \label{eqn:splitting}
  \delta E =& \sqrt{\frac{8}{\pi}}\frac{\mathcal{N}_1^2}{m}
  \left(\frac{\lambda^2}{1+\lambda^2}\right)^{1/4}
  \!\frac{Y(kR)}{\sqrt{kR}}\exp\left(-\!\frac{R}{\xi}\right) \\
Y(kR) =& \cos(kR+\alpha)-\frac{2}{\lambda}\sin(kR+\alpha) \\ &+\frac{2(1+\lambda^2)^{1/4}}{\lambda}, \nonumber
\end{align}
with $\lambda = k \xi$, $2 \alpha = \arctan \lambda$, $k = \sqrt{2 m \mu - \Delta_0^2/v_F^2}$, the Fermi velocity $v_F$ and superconducting coherence length $\xi$. Thus, the effective low-energy model for a multi-vortex configuration reads $H=i\sum_{ij}\delta E_{ij}\gamma_i \gamma_j$, where $\gamma_j$ is a self-conjugate (Majorana) operator representing a zero-energy state in $j$-th vortex. Given that in realistic systems $kR_{ij}\gg 1$, $\delta E_{ij}$ is a rapidly oscillating function.

We now study how this ground-state degeneracy is modified by vortex-antivortex proliferation above the KT transition. We consider a situation where thermal fluctuations affect only the phase of the order parameter while the magnitude remains approximately constant. Below, we will give evidence from a self-consistent calculation that such a regime can be obtained. The fluctuations of the phase can be modeled by a classical XY Hamiltonian for the phases $\theta_{ij}$,
\begin{equation}
H = -J \sum \cos( \arg \theta_{ij} - \arg \theta_{jk} ),
\end{equation}
where $J$ is related to the superfluid stiffness.
%A key property of this model is that excitations are topological defects, namely vortices.
A key property of this model is that below the KT temperature ($T_{\rm KT} = 0.89 J$ in the infinite system\cite{wolff1989-2,gupta1992})vortices and antivortices are bound in pairs by a logarithmic attraction. Above the transition, they unbind and proliferate. The Monte Carlo sampling is performed using a standard cluster update method.\cite{wolff1989,wolff1989-2}

To study the effect of thermal fluctuations on the topological degeneracy, we introduce a fixed vortex/antivortex pair in the system by adding a non-fluctuation phase factor to $\Delta_{ij}$ and study the energy splitting in the presence of the background defects. The fixed vortex-antivortex pair is implemented by introducing an additional phase factor,
\begin{equation} \label{eqn:phase}
\theta_{ij} =  \exp \left( i \phi_{ij} \right) = \exp \left( i \phi_{ij}^A \right) \exp \left( - i \phi_{ij}^B \right).
\end{equation}
Here $\phi_{ij}^A$ ($\phi_{ij}^B$) are the polar angles that the bond $ij$ has with the vortex (antivortex) located at position $A$ ($B$). When applying this to a torus mapped to a lattice with periodic boundary conditions, special care has to be taken that the order parameter is smooth around the boundary.

% long version of the same paragraph
To obtain a simpler description on the torus, we perform a gauge transformation after which the vortices are implemented only by a $\pi$ phase shift in both the hopping and the pairing terms across a particular line (branch cut) connecting the two vortices.
To this end, we introduce a gauge field $\phi_i$ on the sites such that $\phi_{ij} = (\phi_i + \phi_j)/2$ and perform the gauge transformation $c_i \rightarrow c_i \exp \left( -i \phi_i \right)$. For a single vortex, the field $\phi_i$ would wind from 0 to $\pi$ around the vortex, whereas $\phi_{ij}$ would wind from 0 to $2\pi$; therefore $\exp (i \phi_{ij})$ is smooth, while $\exp (-i \phi_i)$ has a jump from $-1$ to 1 at a branch cut. It follows that everywhere except along this branch cut, $\exp( -i \phi_i ) \exp( -i \phi_j ) \exp (- i \phi_{ij} ) = 1$ and the phase is removed from the anomalous hopping; along this branch cut, a phase $\pi$ remains and hopping terms with $i$ and $j$ on different sides of the branch cut pick up a minus sign. For the situation of two vortices, this can be generalized and one finds that the branch cut turns into a line connecting the two vortices. Applying the gauge transformation in the normal hopping, one finds that the same phase factor across the branch cut is introduced.

\section{Results}

\begin{figure}
  \includegraphics{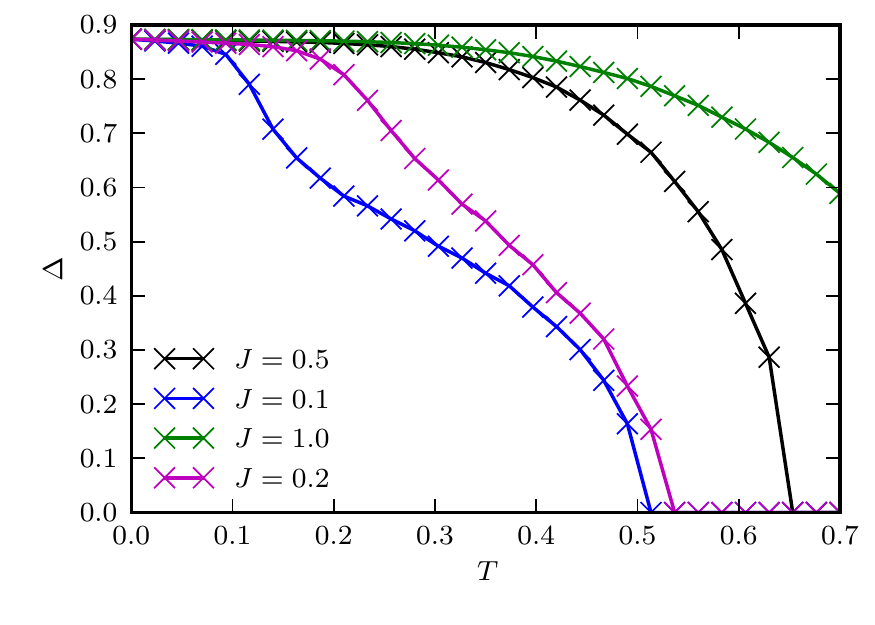}
  \caption{(color online) Self-consistent calculation for $\Delta$ for lattice of size $L=16$ and $U=5$. The Kosterlitz-Thouless transition takes place at $T/J = 0.89$ and leads to the first drop in the gap $\Delta$, which renormalizes it quantitatively while the system remains in a superconducting phase. For even higher temperatures, the gap vanishes and the superconductivity is destroyed. \label{fig:delta} }
\end{figure}

\subsection{Self-consistent calculation}

Owing to the mean-field approximation, the BdG equations have to be augmented with self-consistency conditions,  which read
\begin{align}
\Delta_{ij} &= U \langle c_i c_j \rangle \\
&= U \sum_{E_n > 0} u_n^* (i) v_n (j) \tanh \left( \frac{E_n}{2T} \right) \\
\Delta_0 &= \langle |\Delta_{ij} | \rangle \text{ (spatial and MC average)}.
\end{align}

We have performed self-consistent calculations for a spinless $p$-wave superconductor without fixed vortices and obtained the order parameter $\Delta = \sqrt{ \langle c_i c_j \rangle ^2 }$ as a function of temperature. Our results are shown in Fig.~\ref{fig:delta} for different values of $J$, which controls the relative temperature scales of the Kosterlitz-Thouless transition and the mean-field transition where superconductivity is destroyed. Our data show that for sufficiently small $J$, these two transitions are well-separated and there is an intermediate regime where the phase of the order parameter is disordered, but its magnitude remains finite at a value 20-30~\% below the zero-temperature result. Since we are interested in qualitative results only, such a small qualitative change is irrelevant and we do not perform a self-consistent calculation but fix a value of $\Delta_0$ independent of temperature.

\begin{figure}
  \includegraphics{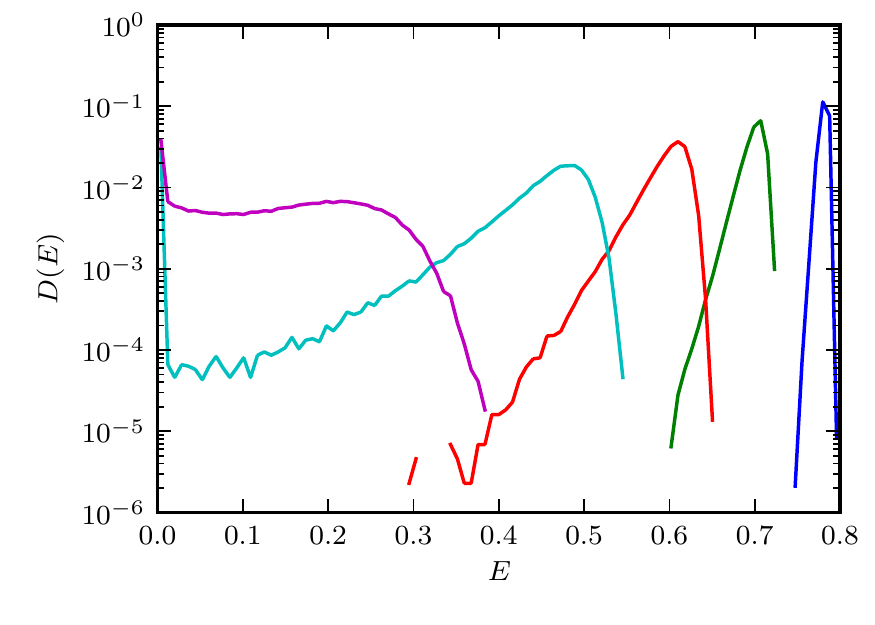}
  \caption{(color online) Density of states $D(E)$ in the low-temperature phase ($T=0.1,0.3,0.5,0.7,0.9$, from right to left) for $L=64$. The DoS is strongly suppressed for sufficiently low temperatures and only shows a peak close to zero energy corresponding to a slightly renormalized value of the energy of the $T=0$ case, and an increase as the transition is approached. Simulations were performed for $\Delta_0=t/2$ and $\mu=-t$. \label{fig:lowT} }
\end{figure}

\subsection{Low-$T$ phase}

Diagonalizing the BdG equation for each configuration of $\theta_{ij}$, we compute the DoS
%By diagonalizing the corresponding BdG equations for each configuration of the field $\theta_{ij}$, we compute the thermal average DoS
\begin{equation}
D(E) = \frac{1}{N} \left \langle \sum_{n} \delta (E - E_n) \right \rangle_T,
\end{equation}
where $\langle \cdot \rangle_T$ indicates the Monte Carlo average at temperature $T$, and $N$ denotes the number of states. We generally average over at least 10,000 configurations and obtain error bars with a standard Jackknife analysis. As shown in Fig.~\ref{fig:lowT}, the DoS at zero temperature shows a sharp peak at the energy splitting set by the system size for the fixed vortices, and a continuum of states above the bulk gap $\Delta_0$. At low temperatures $T\ll T_{\rm KT}$, both features are broadened but the energy splitting of the Majorana modes remains exponential and the DoS is suppressed between this scale and the bulk gap. %This is confirmed by our numerical simulation shown in Fig.~\ref{fig:lowT}.

\begin{figure}
  \includegraphics{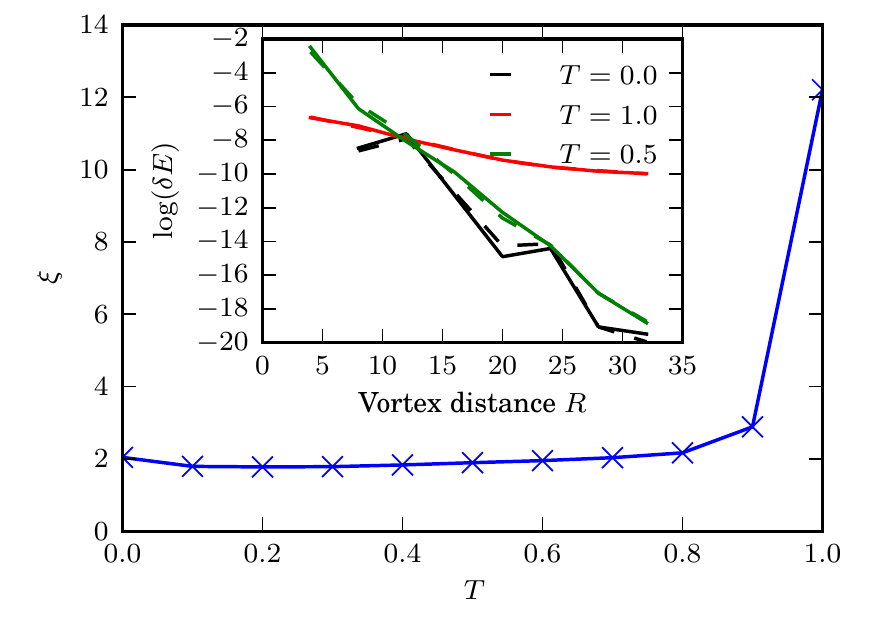}
  \caption{(color online) Scaling of the energy splitting with vortex distance in a system of size $L=64$. The main panel shows the dependence of the coherence length $\xi$ on the temperature. A KT transition takes place at $T \sim 0.9$. The inset shows the dependence of the splitting $\epsilon$ on the distance with a fit to Eqn.~\eqnref{eqn:fit}. Simulations were performed for $\Delta_0=t/2$ and $\mu=-t$. \label{fig:lowT2} }
\end{figure}

To further elucidate the fate of the ground-state degeneracy, we study the energy splitting between fixed vortices by fitting it to (cf. Eqn.~\eqnref{eqn:splitting})
\begin{equation} \label{eqn:fit}
\delta E = \frac{c_1}{\sqrt{R}} \exp \left( -\frac{x}{\xi} \right) \left( 1 + c_2 \cos ( c_3 x+c_4 ) \right),
\end{equation}
where $\xi$, $c_1$, $c_2$, $c_3$ and $c_4$ are fit parameters. Good fits are obtained for low temperatures, as shown in the inset of Fig.~\ref{fig:lowT2}. Our results for the correlation length are shown in the main panel of that figure. The correlation length depends only weakly on temperature as long as the system is well below the KT transition. At the transition, a sharp jump in the correlation length indicates a fundamental change in the scaling behavior.

\subsection{High-$T$ phase}

\begin{figure}
  \includegraphics{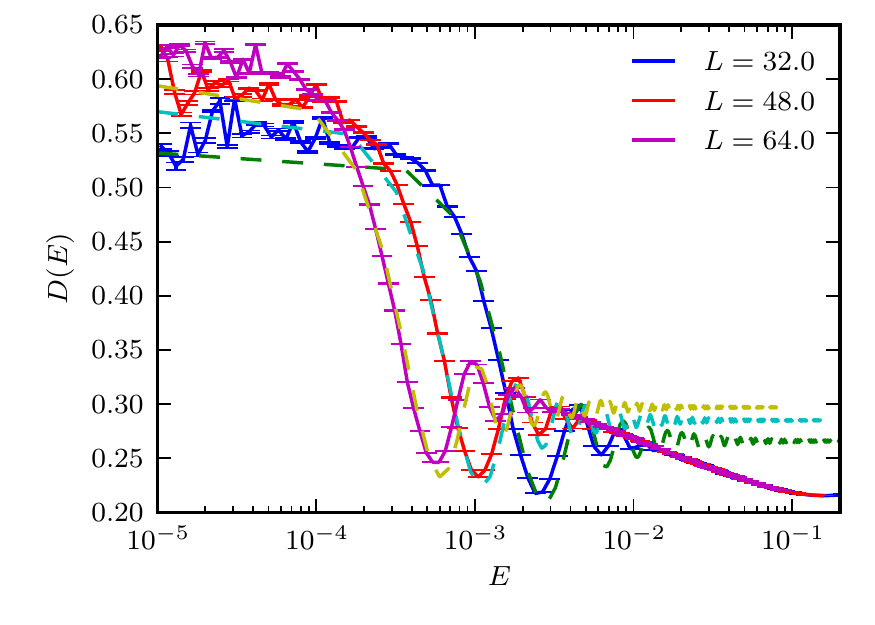}
  \caption{Density of states $D(E)$ in the high-temperature phase ($T=1.5$, $\Delta_0 = t/2$, $\mu = -t$). Fits are to Eqn.~\eqnref{eqn:rmt_D}. The data has been rescaled such that $D(0.03) = 1$. \label{fig:highT} }
\end{figure}

The sharp change is related to the delocalization transition ({\it i.e.} appearance of a disorder-driven thermal metal phase) characteristic to class $D$ superconductors.\cite{bocquet2000,chalker2001,mildenberger2007, laumann2012} The TM is characterized by delocalized states at $E=0$ and a logarithmic divergence of the DoS for low energies.\cite{Senthil00, laumann2012} Furthermore, the oscillatory behavior of the DoS in the zero-dimensional limit is consistent with the random matrix theory predictions for class D.\cite{altland1997} In Fig.~\ref{fig:highT}, the DoS for a spinless $p$-wave superconductor well above the KT transition is shown along with a fit to the random matrix theory result\cite{altland1997} for symmetry class $D$
\begin{equation} \label{eqn:rmt_D}
D(E) \sim \gamma + \frac{ \sin ( 2 \pi \gamma E L^2) }{2 \pi E L^2}.
\end{equation}
Using a single-parameter fit, we obtain excellent agreement with our theoretical expectations for the TM phase: i) At the lowest energies, the DoS follows random matrix theory predictions. ii) For higher energies (but still well below the bulk gap $\Delta$), a logarithmic divergence is observed. This clearly establishes that there is a TM phase above the KT transition, as shown along the $B_y=0$ line in Fig.~\ref{fig:pd}.

The coefficient $\gamma$ above is related to the effective bandwidth in the Majorana fermion hopping problem defined above and should therefore be related to $\delta E$~\eqnref{eqn:splitting}. Indeed, we numerically confirm that $\gamma \sim \Delta_0 \exp ( -\Delta_0/C )$
for some constant $C$. The energy scale is largely independent of temperature as long as the temperature is sufficiently far away from the KT transition.

%We now consider perturbations that break SU(2) symmetry. HQVs correspond to a phase twist only in one spin sector and so still carry Majorana zero-energy modes. However, FQVs do not carry robust zero-energy modes. 

\begin{figure}
  \includegraphics{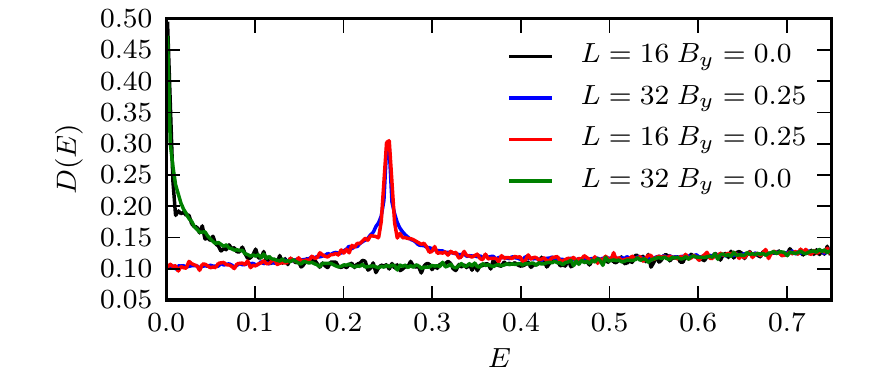}
  \includegraphics{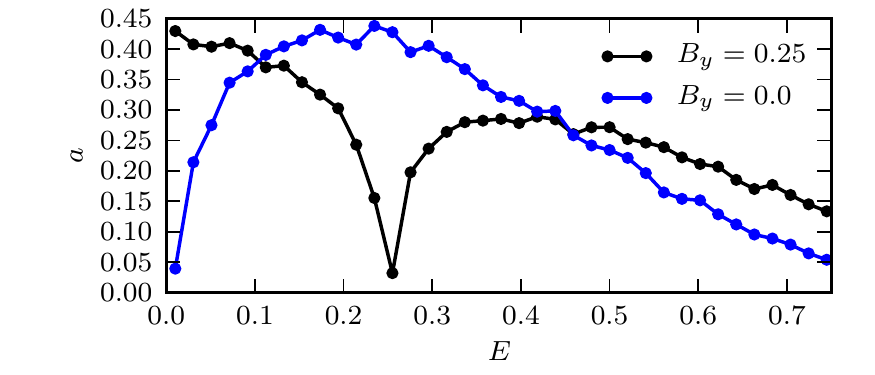}
  \caption{(color online) Top panel: Density of states for spinful fermions. With magnetic field, the DoS has a constant value at $E=0$, whereas without magnetic field it diverges due to the thermal metal. In the case with magnetic field, a peak appears at $E = B_y$. Bottom panel: Constant term obtained from a fit to $I \sim a+L^b$. Values of $a \rightarrow  0$ and $a\neq 0$ indicate extended and localized states, respectively. \label{fig:spinful} }
\end{figure}

\subsection{Spinful model}

When adding perturbations that break SU(2) symmetry, HQVs still carry zero-energy modes, but FQVs do not.
For example, a Zeeman splitting generated by an in-plane magnetic field  $B_y \sum_i \left( i c_{i \uparrow}^\dagger  c_{i \downarrow} - i c_{i \downarrow}^\dagger c_{i \uparrow} \right)$ will move the the lowest excitation energy supported by an FQV to non-zero $E_0= B_y$.
Indeed for each pair of wavefunctions of the spinless model at energies $\pm E$, there are four wavefunctions at energies $\pm E\pm B_y$ for this special choice of field direction.
Thus, if the system has a band of delocalized states near $E=0$, the system will remain in a TM phase for $B_y$ smaller than the width of this band and will transition to an insulating phase once $B_y$ is larger than the width of this band.  To determine this width, in Fig.~\ref{fig:spinful} we show the DoS and the localization properties of the states, which we characterize by the inverse participation ratio (IPR) $I(E)$ defined to be the average of the {\it fourth} moment of a wavefunction of energy $E$.
The inverse participation ratio can be calculated in our setup using
\begin{equation}
I(E) = \left \langle \sum_n \frac{ \langle u_n \rangle ^4 + \langle v_n \rangle ^4 }{ (\langle u_n \rangle ^2 + \langle v_n \rangle ^2)^2 } \delta(E - E_n) \right \rangle_T.
\end{equation}
For finite systems, the IPR must be calculated by averaging over states in a finite range of energies centered around $E$ by broadening the $\delta$ function.

For extended states, the dominant scaling of the IPR is expected to be $I(E) \sim L^{2-\nu}$, where $\nu$ is a non-universal correction to the exponent~\cite{wegner1980,*falko1995,*falko1995-1} while the IPR is expected to approach a constant for localized states, $\lim_{L \rightarrow \infty} I(E) > 0$. Therefore, in the thermal metal phase of the spinless model (or equivalently the spinful model at $B_y=0$), we expect $I(E)$ to scale with a power law at zero energy due to the presence of extended states. At non-zero energies, it is expected to approach a constant value for $L \rightarrow \infty$. 
In the spinful case at finite magnetic field, we expect delocalized states, and hence a power-law scaling of the IPR, at $E \sim B_y$, while we expect localized states and a saturation of the IPR for other energies $E < \Delta$.
In all cases, we expect it to behave with a power law for energies higher than the local gap $\Delta$.

\begin{figure}
  \includegraphics{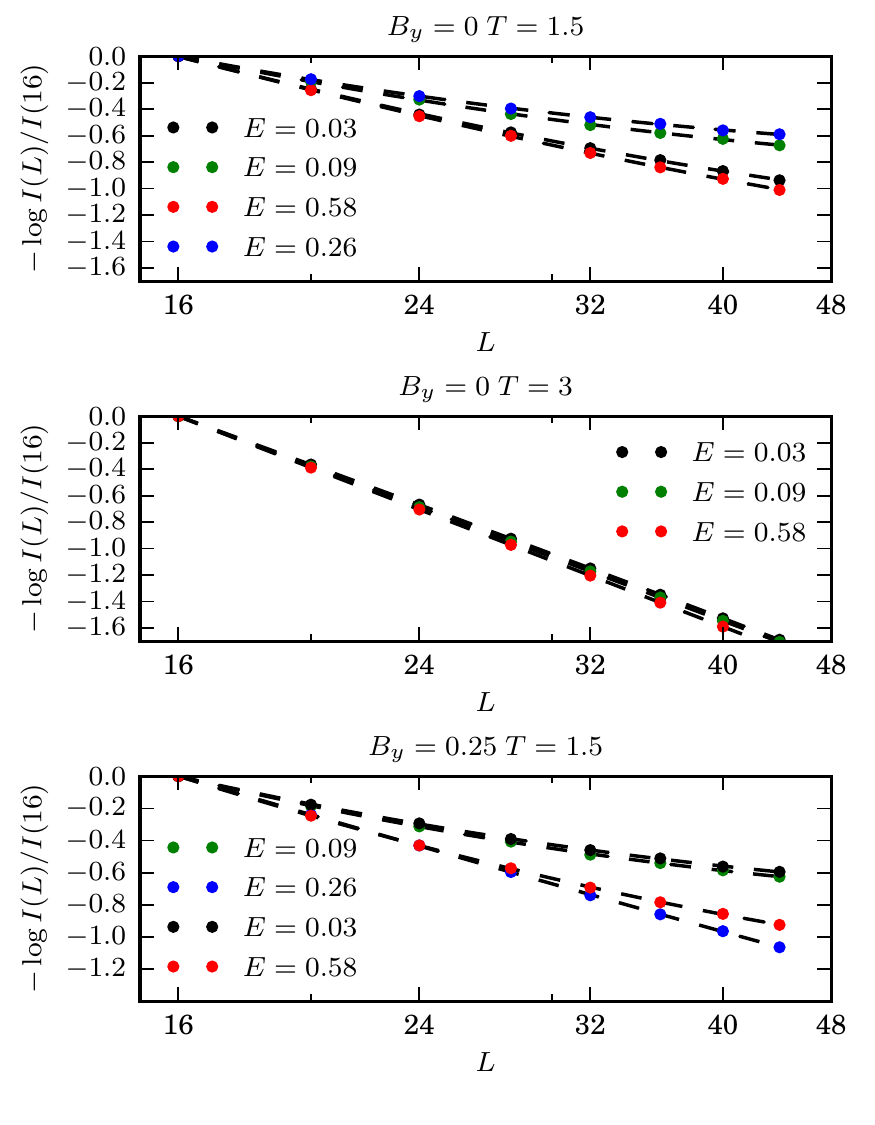}
  \caption{(color online) Fit of the IPR to $I \sim a+L^b$ for several parameters sets. The top panel shows $B_y=0$, $T=1.5$; the center panel shows $B_y=0$, $T=3$, and the bottom panel shows $B_y=0.25$, $T=1.5$. At $T=1.5$, there is clearly only a very narrow band of delocalized states around $E \sim B_y$. For $T=3$, however, the band is broadened such that states at almost all energies appear delocalized for the system sizes we can access. \label{fig:part3} }
\end{figure}

In the bottom panel of Fig.~\ref{fig:spinful}, our results obtained from the extrapolation of the IPR are summarized. These results indicate the presence of delocalized states at energy $E \sim B_y$ and $E > \Delta$, as expected.
A more detailed perspective is provided in Figure~\ref{fig:part3}, which shows several fits for the IPR for the spinful model with two different values of the magnetic field, $B_y = 0$ and $B_y = 0.25$, and two different temperatures $T=1.5$ and $T=3$. The constant terms extract from such fits for $T=1.5$ are shown in Fig.~\ref{fig:spinful}. The top panel of Fig.~\ref{fig:part3} ($B_y=0$, $T=1.5$) clearly shows the saturation of the IPR for energies $0 < E < \Delta$, whereas for very small energies, such as $E=0.03$, no clear sign of saturation is observed for the accessible system sizes. The middle panel shows the same situation for a higher temperature, $T=3$. In this case, the IPR appears to follow a power law also for intermediate energies such as $E=0.09$. This is indicative of a delocalized band of finite width centered around $E=0$, with the bandwidth growing as $T$ is increased. Finally, the bottom panel shows the situation with finite magnetic field and temperature close to the KT transition, where no localization is observed around $E = B_y$, i.e. the IPR scales with a power law.

We have checked that our result is robust against other perturbations, such as an additional magnetic field $B_x$. Since our results are restricted to finite-size systems, it remains open whether these states are truly extended in the thermodynamic limit and how this can be connected to theoretical work.

\begin{figure}
  \includegraphics{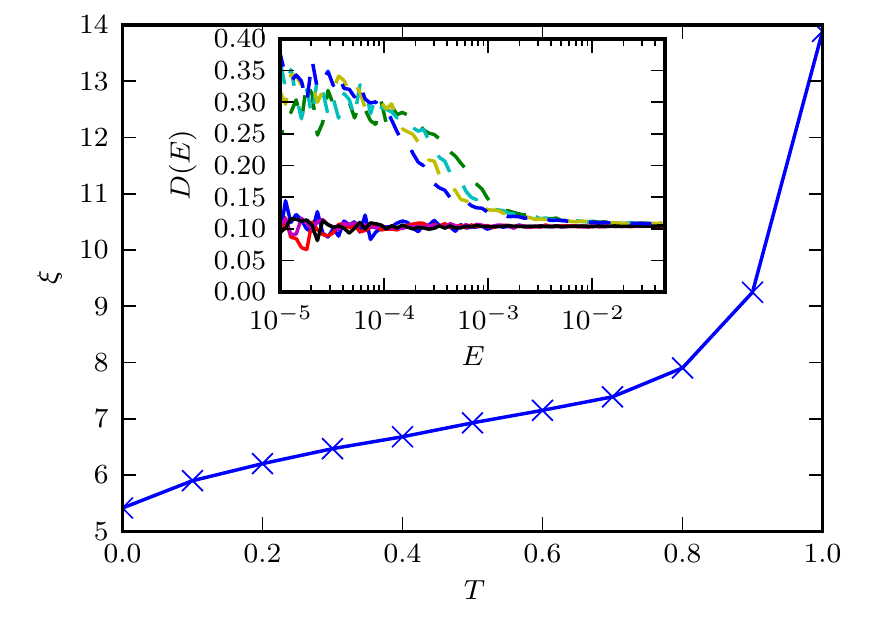}
  \caption{(color online) Main panel: correlation length $\xi$ for HQVs extracted from an analogous fit to Eqn.~\eqnref{eqn:fit}. The correlation length displays a clear jump at the KT transition. Inset: DoS for a system in the thermal insulator phase ($B_y=0.25$, $\Delta=t$, $\mu=-2t$, $L=40,48,56,64$, $T=1.5$) with (dashed lines) and without (solid lines) fixed HQVs. \label{fig:hqv} }
\end{figure}

\subsection{Half-quantum vortices}

We now consider the Majorana zero-energy modes carried by half-quantum vortices (HQVs) in the trivial thermal insulator (TTI) phase by studying the DoS with two fixed half-quantum vortices in the background of thermally fluctuating full quantum vortices. As shown in the inset of Fig.~\ref{fig:hqv}, the HQVs give an additional contribution to the DoS at low energies. Studying the energy splitting as a function of the temperature, we find that the correlation length $\xi$ changes qualitatively at the KT transition, see Fig.~\ref{fig:hqv}.  While our small system sizes do not let us determine whether the splitting is still exponential or becomes power law in this regime, the splitting energy for HQVs clearly changes dramatically above the KT transition.

\begin{figure}
  \includegraphics{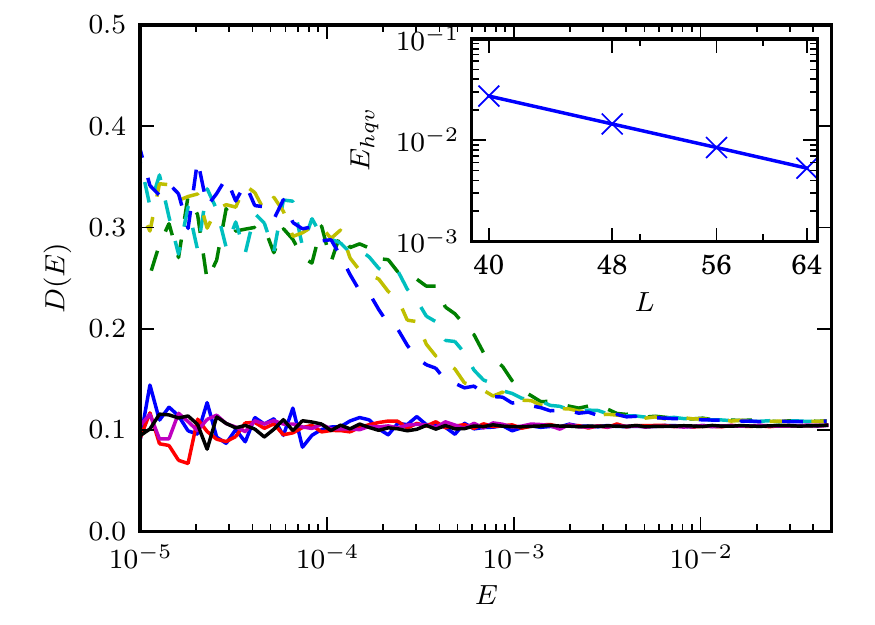}
  \caption{Contribution of two fixed half-quantum vortices to the density of states in the high-temperature phase of a spinful superconductor with finite magnetic field. The main panel shows the density of states (solid lines: with HQVs, dashed lines: without HQVs). The scaling of the characteristic energy scale $E_{hqv}$, defined below Eqn.~\eqnref{eqn:integr_diff}, is shown in the inset on a log-log scale. \label{fig:hqv2} }
\end{figure}

As an alternative approach to quantify the energy scale below which half-quantum vortices contribute, we study the integrated difference between the density of states with, $D^h(E)$, and without, $D(E)$, half-quantum vortices:
\begin{equation} \label{eqn:integr_diff}
\rho(E) = \int_0^E dE' \left( D^h(E') - D(E') \right)
\end{equation}
and define $E_{hqv}$ as the lowest energy such that $N \rho(E_{hqv}) = 1$, where $N = 2L^2$.

Fig.~\ref{fig:hqv2} shows the density of states with and without half-quantum vortices (cf. inset of Fig.~\ref{fig:hqv}). In the inset, the scaling of this quantity with system size is shown. A power-law scaling is clearly observed. This is consistent with the observation that the expectation value of the lowest energy $\langle E_0 \rangle_T$ behaves like a power-law both with and without HQVs. This shows that the topological degeneracy is destroyed by the presence of zero-energy states due to disorder, even though these states are localized.

\section{Conclusions}

We have studied the effect of thermal fluctuations on two-dimensional chiral $p$-wave superconductors in symmetry class $D$. We have shown that thermally disordering the superconducting phase drastically changes the topological properties of these systems. We can explain the underlying mechanism as the proliferation of vortex-antivortex-pairs, which carry low-energy excitations in their cores. Hybridization of these low-energy states gives rise to a thermal metal phase. Exploring the full phase diagram as a function of temperature and in-plane magnetic field, we find that in addition to the low-temperature topological superconductor and the thermal metal phase, a trivial thermal insulator (Anderson insulator) phase appears. We study the fate of the topological degeneracy in all these phases and find that it is well-defined only in the topological superconductor phase at $T < T_{\rm KT}$. In the thermal insulator phase, the splitting of the degeneracy due to half-quantum vortices changes dramatically due to vortex disorder.
This result can be anticipated considering previous analytical evidence for a disorder-driven quantum phase transition in one-dimensional analogues of our system.~\cite{motrunich2001,gruzberg2005}

Throughout this paper, we have assumed $\Delta \sim \mu$ in order to keep the coherence length on the order of a few lattice sites. In experimental systems, however, $\Delta$ is often much smaller than $\mu$. In this case, one expects to find many subgap states localized on each vortex with energy splitting $\varepsilon_s \sim \Delta^2/\mu$.\cite{meng2012} While this regime is difficult to treat numerically, we argue based on our results for the high-$T$ phase in magnetic field that there are two scenarios for the high-$T$ phase in the presence of many subgap states: if the hybridization scale for Majorana modes obeys $\delta E \leq \varepsilon_s$, the subgap states will contribute to the delocalized states in the thermal metal band which is therefore enhanced. If, on the other hand, $\varepsilon_s \gg \delta E$, one would expect several peaks in the DoS centered around the subgap state energies $\varepsilon_s \cdot n$ similar to the peak around the magnetic field. Here $n$ is an integer corresponding to $n$-th energy level in the vortex core. In either case, thermal fluctuations destroy the topological phase and our qualitative conclusions regarding the disordered high-$T$ phases remains valid.

\acknowledgements

{\it Note added.} After completing this work, we became aware of Ref.~\onlinecite{nandkishore2012}, which also discusses topological superconductors above the KT temperature, but reaches a different conclusion regarding the nature of the phase.

We acknowledge useful discussions with Matthew Fisher, Andreas Ludwig, Achim Rosch and Simon Trebst. We thank the Aspen Center for Physics where this work was initiated. Simulations were performed using the ALPS libraries.\cite{bauer2011-alps}

\bibliographystyle{apsrev4-1}
\bibliography{refs}

\begin{thebibliography}{44}
\bibitem [{foo()}]{footnote1}%
  \BibitemOpen
  \href@noop {} {}\bibinfo {note} {We assume throughout the paper that the
  system size is smaller than the localization length $L_c$.}\BibitemShut
  {Stop}%
\end{thebibliography}

\end{document}